\documentclass[%
 aip,
 floatfix,
 amsmath,amssymb,
 reprint,%
]{revtex4-1}

\usepackage{graphicx}
\usepackage{dcolumn}
\usepackage{bm}

\usepackage[utf8]{inputenc}
\usepackage[T1]{fontenc}
\usepackage{mathptmx}
\usepackage{etoolbox}

\usepackage{physics}
\usepackage[dvipsnames]{xcolor}
\usepackage{tikz} 
\usepackage{float}
\usepackage[breaklinks=true, hidelinks]{hyperref}

\definecolor{drkgreen}{rgb}{0.0, 0.5, 0.0}
\definecolor{violet}{rgb}{0.5, 0.0, 1.0}

\DeclareUnicodeCharacter{2009}{\,}

\makeatletter
\def\@email#1#2{%
 \endgroup
 \patchcmd{\titleblock@produce}
  {\frontmatter@RRAPformat}
  {\frontmatter@RRAPformat{\produce@RRAP{*#1\href{mailto:#2}{#2}}}\frontmatter@RRAPformat}
  {}{}
}%
\makeatother

\begin{document}

\preprint{APS/123-QED}

\title{Control of multi-modal scattering in a microwave frequency comb}
\author{J. C. Rivera Hernández}
 \email{jcrivera@kth.se}
 \affiliation{Department of Applied Physics, KTH Royal Institute of Technology, SE-10691 Stockholm, Sweden} 
 
\author{Fabio Lingua}
 \affiliation{Department of Applied Physics, KTH Royal Institute of Technology, SE-10691 Stockholm, Sweden}

\author{Shan W. Jolin}
 \affiliation{IQM Quantum Computers, FI-02150 Espoo, Finland}

\author{David B. Haviland}
 \affiliation{Department of Applied Physics, KTH Royal Institute of Technology, SE-10691 Stockholm, Sweden}

\date{\today} 

\begin{abstract}

Control over the coupling between multiple modes of a frequency comb is an important step toward measurement-based quantum computation with a continuous-variable system.
We demonstrate the creation of square-ladder correlation graphs in a microwave comb with 95 modes.
The graphs are engineered through precise control of the relative phase of three pumps applied to a Josephson parametric oscillator.
Experimental measurement of the mode scattering matrix is in good agreement with theoretical predictions based on a linearized equation of motion of the parametric oscillator.
The digital methods used to create and measure the correlations are easily scaled to more modes and more pumps, with the potential to tailor a specific correlation graph topology.

\end{abstract}

\maketitle


\section{Introduction}

The storage and manipulation of information encoded in quantum states of the harmonic oscillator is attracting a lot of interest due to its promise of measurement-based quantum computation (MBQC).
Controlling the infinite-dimensional Hilbert space of this continuous variable (CV) system is motivated by the speed of control, resilience against decoherence and the ability to correct errors~\cite{grimm_stabilization_2020, kudra_robust_2022, sivak_real-time_2023}.
A reduction in complexity, of both the quantum hardware and the classical control and readout hardware, further motivates the development of controlled entanglement in larger CV systems.
Seminal works have paved the way, providing the building blocks of MBQC~\cite{braunstein_quantum_2005, weedbrook_gaussian_2012, pfister_continuous-variable_2019}, and demonstrating the ability to engineer the structure of quantum correlations in the form of graph or cluster states~\cite{menicucci_universal_2006, menicucci_one-way_2008, gu_quantum_2009, menicucci_graphical_2011}.
Multi-frequency mixing processes in non-linear optical systems are typically used to create these correlations, described as combinations of multi-mode squeezing and beam-splitter operations~\cite{pfister_multipartite_2004, bensemhoun_multipartite_2023}.

Multi-modal squeezing at optical frequencies can now be realized on a chip~\cite{yang_squeezed_2021, jahanbozorgi_spectroscopic_2023} and through spectral and temporal multiplexing of ultra-fast pulses~\cite{kouadou_spectrally_2023}.
CV cluster states have been realized in optical frequency combs~\cite{menicucci_arbitrarily_2010, pysher_parallel_2011, cai_multimode_2017, walschaers_tailoring_2018} and with time-multiplexing techniques~\cite{menicucci_temporal-mode_2011, chen_experimental_2014, asavanant_time-domain-multiplexed_2021, du_generation_2023}.
Time multiplexing not only facilitates the realization of CV quantum gates but also supports MBQC protocols~\cite{menicucci_temporal-mode_2011, larsen_deterministic_2019, asavanant_time-domain-multiplexed_2021, du_generation_2023}.
Optical frequencies benefit from the fact that at room temperature the thermal energy $k_BT$ is lower than the photon energy $hf$, but the analog signal processing techniques used in optics lack the flexibility and precision of digital signal processing. 
At optical frequencies it is difficult to easily and independently control the relative phase of each individual pump, an effective tool for engineering correlations~\cite{jing_experimental_2006, hung_quantum_2021, petrovnin_generation_2023, pfister_cluster_2023}.

Recent advances in digital signal processing at microwave frequencies, combined with millikelvin cooling technologies, are enabling new platforms for CV quantum technologies.
Phase control in a multimode superconducting parametric cavity has been used to simulate ladder-like bosonic lattices~\cite{hung_quantum_2021}.
Multi-mode squeezing has been demonstrated with surface acoustic wave resonators~\cite{andersson_squeezing_2022}, Josephson~\cite{naaman_synthesis_2022, petrovnin_microwave_2023}, and traveling-wave parametric amplifiers~\cite{esposito_observation_2022}.
Multi-partite entanglement~\cite{jolin_multipartite_2023} and cluster states~\cite{petrovnin_generation_2023} have been realized with microwave superconducting circuits.

Here we demonstrate the creation of 2D square-ladder correlation graphs of a microwave frequency comb, generated by three phase-coherent pumps driving a single Josephson Parametric Amplifier (JPA).
The circuit diagram in Fig.~\ref{fig:schematics}a describes the schematic of the JPA, corresponding to an LC harmonic oscillator with an effective time-varying inductance controlled by the pump signal.
We reveal the correlations through analysis of the scattering of a coherent state of approximately $7\times 10^3$ photons, realized by digitally synthesized single-tone microwave signal $\sim 4.2$~GHz, properly thermalized at 10 mK.
By controlling the relative phase of one of the three pumps, we selectively modify the output correlations.
Verification of the scattering between input and output modes is an important step towards multi-modal entanglement in a useful form.
Mode scattering theory is an effective tool for predicting entanglement of Gaussian states~\cite{weedbrook_gaussian_2012, ranzani_graph-based_2015, serafini_quantum_2017, peterson_parametric_2020, naaman_synthesis_2022}.
We use this theory to predict scattering matrices that we verify through experiments.

\section{Theoretical Framework} \label{teo}

The parametrically driven JPA is described by a time-dependent quadratic Hamiltonian~\cite{yamamoto_principles_2016}
\begin{equation}
    H=\frac{\omega_0}{2}(A^\dag A + AA^\dag) + \frac{\omega_0}{2}g_p(t)[A + A^\dag]^2,
    \label{H0}
\end{equation}
where $A$ ($A^\dag$) is the annihilation (creation) operator of the time-independent Hamiltonian of the harmonic oscillator with resonant frequency $\omega_0$, and the arbitrary pump waveform $g_p(t)=\sum_k A_k \cos{(\Omega_k t +\phi_k)}$ is composed of multiple tones with different frequency, amplitude and phase.

\begin{figure}
\includegraphics[width=\columnwidth]{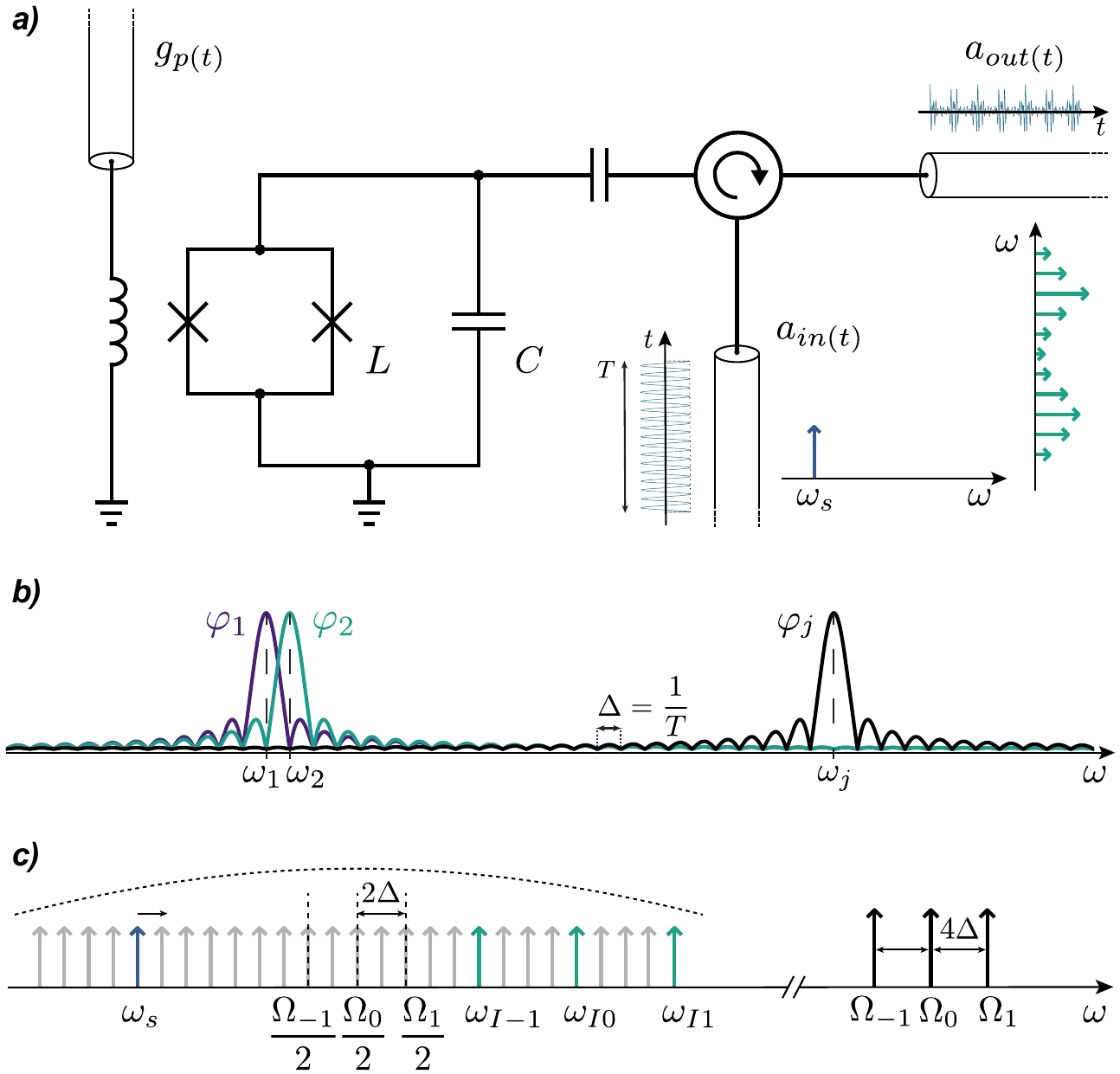}
    \caption{\label{fig:schematics} a) Schematic of the scattering experiment. 
    An external time-dependent flux $g_p(t)$ modulates the JPA's inductance, enabling frequency mixing, also called intermodulation. 
    A coherent state (i.e. a single-tone microwave signal at frequency $\omega_s \approx 2 \pi \times 4.2$~GHz, properly thermalized at $T=10$~mK) injected into the input port mixes with the pumps, generating several intermodulation products which are measured at the output port.
    b) The measurement basis in the frequency domain. 
    Fourier leakage is eliminated and mode orthogonality is established by tuning each mode frequency $\omega_j$ to fall on the \emph{sinc} nodes of the other modes $\omega_i$, $i\neq j$, satisfied when the frequency spacing is equal to the measurement bandwidth $\Delta=1/T$. 
    c) Illustration of the pumping scheme. 
    The flux pump with three tones (black) at $\Omega_{k}=2\omega_0 + 4k\Delta$, with $k=\{-1,0,1\}$.
    The microwave signal (blue) is stepped through all the modes (grey) generating 2nd and 3rd-order intermodulation products symmetrically around $\Omega_{k}/2$. 
    For the sake of clarity we draw only 2nd order products (green).}
\end{figure}

Both classical and quantum-canonical treatments of a quadratic Hamiltonian lead to the same equation of motion, a Langevin-like equation describing the field quadratures inside the JPA~\cite{yamamoto_principles_2016}
\begin{equation}
    \frac{dA}{dt}=-i\tilde\omega_0A-i\omega_0g_p(t)(A + A^\dag) +\sqrt{\gamma}\;a_{\text{in}}.
    \label{EOM0}
\end{equation}
Here $a_{\text{in}}$ is an external signal input through a transmission line, $\tilde\omega_0=\omega_0 - i\frac{\gamma}{2}$, and $\gamma$ the effective coupling between the transmission line and the JPA~\cite{yamamoto_principles_2016, peterson_parametric_2020}. 
In this work we analyze the dynamics in the classical regime where the operators $A$ ($A^\dag$) can be treated as complex numbers $A$ ($A^*$).

We design an orthogonal basis of modes for measurement and control, defined from the duration $T$ of sequential measurement time windows, $c_j(t)=e^{-i\omega_jt}/\sqrt{T}$ with $\omega_j= 2\pi/T_j$, with mode period $T_j$ commensurate to $T$ (see Appendix~\ref{appA} for details).
In the frequency domain each mode $j$ is represented by a \emph{sinc function} $\varphi_j\equiv\varphi(\omega-\omega_j)$, equally spaced by $\Delta$ and centered at $\omega_j$, as shown in Fig.~\ref{fig:schematics}b.
With this choice of basis, the maximum of each mode is at the nodes of all the other modes, ensuring orthogonality and preventing leakage between modes.

Representing the field quadratures of Eq.~(\ref{EOM0}) in this mode basis $A=\sum_ia_i(t)=\sum_ia_ic_i(t)$ $\left(A^\dag=\sum_ia_i^*c_i^*(t)\right)$, one obtains a set of $n$ coupled, linear harmonic balance equations
\begin{equation}
    \left\{\begin{array}{rcl}
        i\omega_ia_i + i\tilde\omega_0a_i +i\sum_j g_{ij}(a_j + a_j^*)&=&\sqrt{\gamma}\;a_{\text{in},\,i}\\
        -i\omega_ia_i^* - i\tilde\omega_0^*a_i^* -i\sum_j g_{ij}(a_j + a_j^*)&=& \sqrt{\gamma}\;a_{\text{in},\,i}^*
    \end{array}\right. ,\;\;\;\forall i
    \label{eq:nEOM}
\end{equation}
with the terms $g_{ij}$ that couple mode $i$ to mode $j$ determined by the amplitude and phase of the pumps (see appendix~\ref{appA} for details).
The system \eqref{eq:nEOM} can be expressed in matrix form with the usual formalism of mode-scattering theory~\cite{peterson_parametric_2020, naaman_synthesis_2022, petrovnin_generation_2023, jolin_multipartite_2023}
\begin{equation}
    -iM\vec{a}=K\vec{a}_{\text{in}}. 
    \label{eq:matr_form}
\end{equation}
The solutions to the multi-modal set of equations of motion \eqref{eq:matr_form} are readily available upon inversion of the matrix $M$.
Here the mode-coupling matrix $K$ is a diagonal matrix containing the couplings of each mode to the transmission line, assumed to be constant $K_{ii}=\sqrt{\gamma}$.
The components of the vectors $\vec{a}$ and $\vec{a}_{\text{in}}$ in \eqref{eq:matr_form} contain the complex mode amplitudes as $\vec{a}=\left(a_1,a_1^*,a_2,a_2^*,\dots,a_i,a_i^*,\dots\right)^T$.
The matching condition between the input/output and internal modes, $\vec{a}_{\text{in}}+\vec{a}_{\text{out}}=K\vec{a}$, leads to the scattering matrix,
\begin{equation}
    S=iKM^{-1}K-I.
    \label{Snum}
\end{equation}
We numerically compute the scattering matrix through Eq.~(\ref{Snum}) and compare it to that measured in the experiment.

\begin{figure}[h]
\includegraphics[width=\columnwidth]{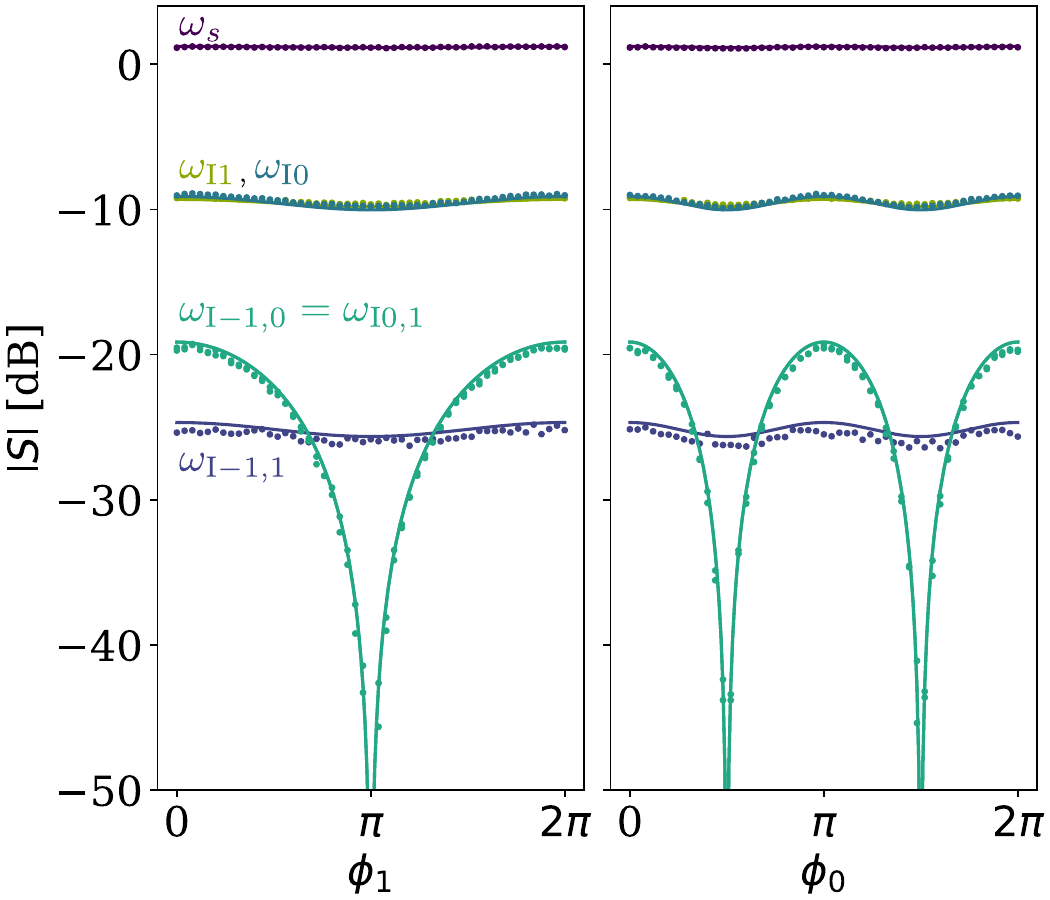}
    \caption{\label{fig:phase-dep} Magnitude of the second and third-order intermodulation products as a function of the phase $\phi_1$ of the edge pump at $\Omega_1$, and phase $\phi_0$ of the central pump at $\Omega_0$. 
    The phase dependence of the other edge pump at $\Omega_{-1}$ (not shown) is identical to that at $\Omega_{1}$. 
    Second-order intermodulation products at $\omega_{\mathrm{I}k}=\Omega_{k}-\omega_s$, $k=\{-1,0,1\}$, show weak variation with pump phase.
    Third-order products at $\omega_{\mathrm{I}k,m}=\Omega_{k}-\Omega_{m}+\omega_s$ ($k \ne m$) show a strong dependence on pump phase, due to interference in the mixing process involving two pumps.  
    The experiment (dots) is in good agreement with the theory (solid lines).  
    For clarity, we only plot some intermodulation products. Identical phase dependence is observed for: $\omega_{\mathrm{I}-1}$ and $\omega_{\mathrm{I}1}$; $\omega_{\mathrm{I}1,-1}$ and $\omega_{\mathrm{I}-1,1}$; $\omega_{\mathrm{I}0,-1}$ with $\omega_{\mathrm{I}-1,0}$.}
\end{figure}

\begin{figure*}[t]
\centering
\includegraphics[width=\textwidth]{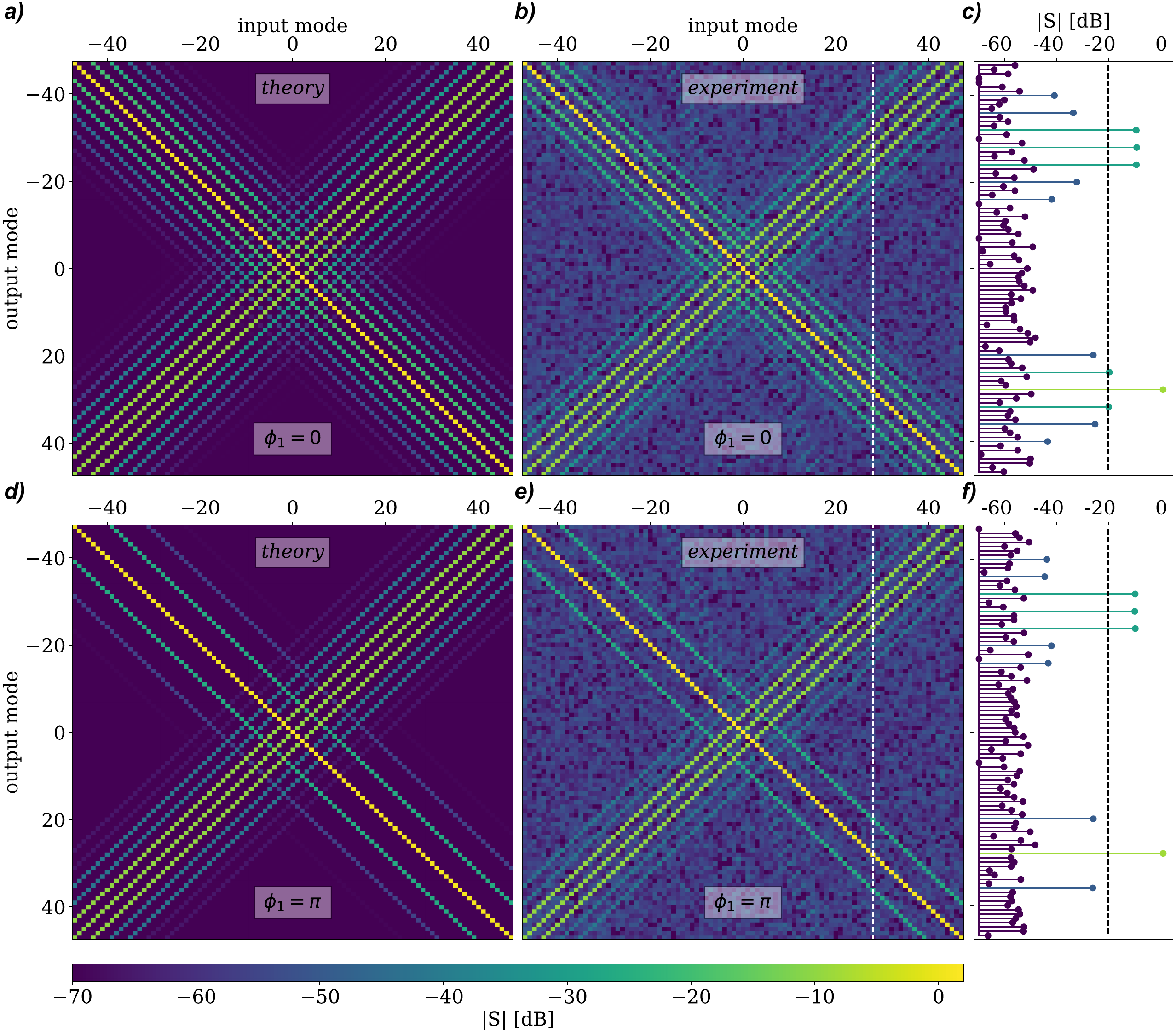}
    \caption{\label{fig:S_3pmp} Theoretical a) and d) and experimental b) and e) scattering matrices of a tripled-pumped JPA. 
    The upper (lower) panels correspond to pump phase $\phi_{1}=0$ ($\phi_{1}=\pi$). 
    We measure 95 modes where mode index $0$ is at half of the central pump frequency, $\omega_0 = \Omega_0/2$. 
    The magnitude of the scattering is expressed in dB, normalized to the pump-off case. 
    c) and f) show one column of the experimental scattering matrix, marked by the white dashed line in b) and e).  
    The black-dashed marks the $-20~\text{dB}$ threshold used to generate the correlation graphs of Fig.~\ref{fig:graphs}.}
\end{figure*}

\section{Experiment and Discussion} \label{exp}

The experiment measures the scattering of a weak input signal that intermodulates with a multifrequency pump inside a Josephson Parametric amplifier, generating multiple intermodulation products, also known as idlers, which leak out of the JPA and into the transmission line.
The JPA acts as a parametric oscillator, a lumped-element LC circuit with L being the Josephson inductance of a SQUID (see Fig.~\ref{fig:schematics}a).
Flux in the SQUID loop, controlled through a separate pump port, modulates the microwave oscillator's inductance or resonant frequency.
In the absence of flux the measured resonance frequency of the JPA is 7.8 GHz, and with $\Phi_{dc} \approx \pm 0.42 \Phi_0$ we measure $\omega_0/2\pi \approx 4.2$ GHz.
At this DC flux bias we apply microwave pumps at frequencies $\Omega_k \sim 2\omega_0$, enabling 3 and 4-wave mixing of signal and pumps.
The signal port is over-coupled to the JPA, giving a parametric oscillator with loaded quality factor $Q=37.5$, corresponding to a linewidth $\kappa = 2 \pi \cross 112$ MHz.
A circulator separates the incoming and outgoing modes, with the latter directed to a cryogenic low-noise amplifier through a double isolator.
A complete schematic diagram of the measurement set-up is given in appendix~\ref{app-setup}.

The pumps are generated and data are collected with a digital multi-frequency lock-in that coherently modulates and demodulates at up to 192 frequencies simultaneously\cite{tholen_measurement_2022}.
The frequencies are tuned to be integer multiples of the measurement bandwidth $\Delta = 1/T$, fixing the frequency resolution and defining the orthonormal set of modes $\{a_i\}$.
In the experiment described here we choose $\Delta = 0.1~\text{MHz}  \ll \kappa$ and we perform IQ demodulation on a basis of 95 tones spanning 0.95~MHz, all within the bandwidth of the JPA.
We begin with the analysis of scattering with one and two pumps, where we observe that both the phase of the single pump and the relative phases between pumps, do not influence the measured scattering matrix (see Appendix~\ref{App12pmp} for details).
Control of the relative phase of three pumps is the key to canceling correlations between modes.
We set three pumps of equal strength at frequencies close to twice the resonant frequency of the JPA, $\Omega_{k} \simeq 2\omega_0+4k\Delta$, with $k=\{-1, 0, 1\}$ and $\Delta << \kappa$.
The central $k=0$ pump and $\Delta$ determine the enumeration scheme for the frequencies in the scattering matrix, $\omega_j = \Omega_0/2 + j \Delta$, with $j \in \mathbb{Z}$.

\begin{figure}
\includegraphics[width=0.95\columnwidth]{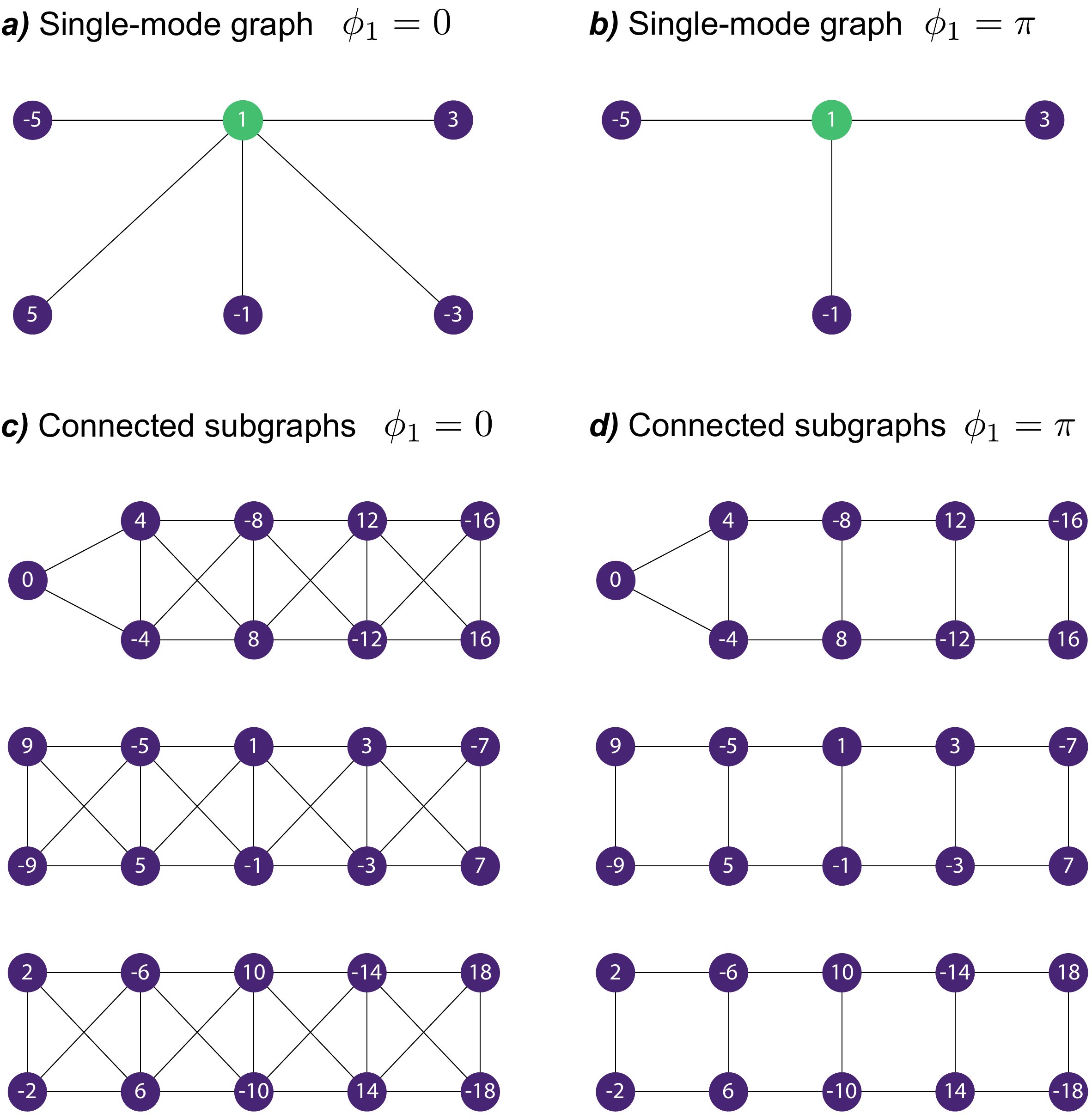}
    \caption{\label{fig:graphs} Single mode correlation graphs at threshold $-20~\text{dB}$ for a) pump phase $\phi_{1}=0$ and b)  $\phi_{1}=\pi$.
    c) and d) show the set of three connected subgraphs obtained by considering correlations among all modes. 
    Pump phase $\phi_1=\pi$ causes destructive interference which removes selected edges to create a square-ladder topology.}
\end{figure}

We inject a signal at $\omega_s$, stepping it through the different $\omega_j$ while listening simultaneously at all 95 frequencies to determine where the signal scatters, as illustrated in Fig.~\ref{fig:schematics}c.
The signal undergoes 3-wave mixing with each pump resulting in the generation of multiple idlers (2nd-order intermodulation products) at frequencies  $\omega_{\mathrm{I}k}=\Omega_{k}-\omega_s$.
The signal also undergoes 4-wave mixing with two pumps to generate 3rd-order intermodulation products at frequencies $\omega_{\mathrm{I}k,m}=\Omega_{k}-\omega_{\mathrm{I}m}=\Omega_{k}-\Omega_{m}+\omega_s$ (for $k \ne m$). 
Fig.~\ref{fig:phase-dep} shows the magnitude of 2nd and 3rd-order intermodulation products as a function of pump phase, or phase of one pump with respect to that of the other two.
In our specific configuration 3rd-order intermodulation products involving different pumps coincide in frequency, leading to constructive or destructive interference, depending upon the pump phase. 
Consequently, manipulation of the phase of one pump allows us to selectively cancel specific products, and thereby engineer correlations in the output modes. 
Rotation of the phase of $k=\pm1$ edge pumps generates $2\pi$-periodic interference (see fig.~\ref{fig:graphs}a) whereas rotation of the central $k=0$ edge pump demonstrates $\pi$-periodic interference (see fig.~\ref{fig:graphs}b). 
The theoretical model is in good agreement with the measurements.

Figure~\ref{fig:S_3pmp} shows the scattering matrices, normalized to the zero-pump case, under conditions of constructive and destructive interference. 
Again, the theoretical scattering matrices show good agreement with the experiment, with the latter limited in resolution by the added noise of the first-stage amplifier.
In fitting the theory we minimize the sum of square deviations between experimental and theoretical scattering matrices adjusting two parameters: the common pump strength (mode coupling) $g$ and an effective transmission line coupling $\gamma$, assumed the same for all frequencies (see Appendix~\ref{app-fit}).

The statistical properties of Gaussian states are fully determined by the first and second moments~\cite{weedbrook_gaussian_2012}.
Correlations between Gaussian modes at different frequencies are therefore defined by the covariance matrix $V$.
When expressed in the frequency-modes basis, the non-zero off-diagonal entries $V_{ij}$, give information about the correlation between the quadratures of modes $i$ and $j$.
One can use the scattering matrix to predict the output correlation between a set of uncorrelated Gaussian modes at the input of the JPA.
For a set of uncorrelated input fluctuations the covariance matrix is proportional to the identity matrix $V_{in} \propto \mathbb{I}$.
The covariance matrix at the output is therefore totally determined by $S$: $V_{out}=S_x V_{in} S_x^T$, where $S_x=USU^\dag$ is the scattering matrix expressed in the quadratures basis (see appendix~\ref{app-covmat} for details).
Furthermore, for the pumping schemes analyzed in this paper, the scattering matrix $S$ and covariance matrix $V$ have the same mode connectivity (i.e. $S_{ij}\neq 0$ implies $V_{ij}\neq 0$).

Through analysis of the scattering matrix, we gain insight into the structure of correlations. 
Each column of the $S$ matrix is proportional to the strength of correlation between that mode and all connected modes in the matrix (see Fig~\ref{fig:S_3pmp}c and f).
To represent the connections we construct single-mode correlation graphs wherein each vertex corresponds to a mode, and edges represent any connection between modes that is above a certain threshold. 
These single-mode graphs are shown in Figs.~\ref{fig:graphs}a and \ref{fig:graphs}b for the cases of constructive and destructive interference shown in Fig.~\ref{fig:phase-dep}a.
Connecting these single-mode graphs we can visualize the correlation topology, as shown in Figs.~\ref{fig:graphs}c and \ref{fig:graphs}d.

Interestingly one can create three independent correlation graphs isomorphic to a nearest-neighbor square ladder when the threshold is set to  $-20~\text{dB}$ below the pump-off level.
This threshold corresponds roughly to the noise floor of our measurement setup when the injected signal is at the level of the vacuum fluctuations.
Lowering the threshold an additional 6~dB would result in a square-ladder graph with next-nearest-neighbor connections, e.g. mode $1$ in Fig.~\ref{fig:graphs}b-d being weakly connected also with mode $9$ and mode $-7$. 
The creation of a square-lattice topology of Fig.~\ref{fig:graphs}d is interesting at the quantum level as it suggests the presence of correlations featuring the canonical graph of a CV cluster state~\cite{menicucci_graphical_2011, pfister_continuous-variable_2019}.
By injecting quantum fluctuations as input of the JPA, one expects to obtain a multi-modal squeezed vacuum at the output mirroring this topology of correlation.

\section{Conclusion}\label{Concl}

Digital signal processing at microwave frequencies was used to engineer mode correlations in a continuous-variable system.
The synchronous excitation and detection of response at multiple frequencies, all derived from one clock and tuned to one base frequency, established a set of orthogonal modes and a global phase reference for control and measurement.
Controlling the relative phases of three pumps driving a Josephson parametric amplifier, we generated the scattering matrix required to create a square-ladder correlation graph.
Creating this graph on a basis of 95 modes is a significant milestone toward manipulating the quantum correlations of a cluster state sufficient for measurement-based quantum computation.
Future work intends to explore and verify entanglement on a square-ladder graph.

The three-pump case represents the simplest non-trivial case where interference can be exploited to affect the topology of the correlation graph.
The addition of many more pumps, with precise control of the amplitude and phase of each, is a trivial extension with the digital microwave platform.
However, the complexity of the scattering problem grows rapidly with the number of pumps.
In this situation the simulated scattering matrices become a useful tool for exploring more complicated pumping schemes.
With precise control of both the phase and amplitude of 95 pump tones with our digital signal processing platform, we can rapidly check that the scattering matrix has the desired graph topology.
These advances provide hope for scaling continuous variable entanglement to larger systems, thereby enabling measurement-based quantum computing.

\begin{acknowledgments}
We acknowledge NIST and J. Aumentado for providing the JPA used in this experiment.
We thank P. Hakonen, K. Petrovnin and E. Mukhanova for fruitful discussions.
This work was supported by the Knut and Alice Wallenberg Foundation through the Wallenberg Center for Quantum Technology (WACQT).
\end{acknowledgments}

\section*{Author declaration}

\subsection*{Conflict of interest statement}
D. B. H. is part owner of the company Intermodulation Products AB, which produces the digital microwave platform used in this experiment.

\subsection*{Author contributions}

\begin{table}[h]
{\footnotesize
\begin{tabular}{l|l|l|l|l|}
\cline{2-5}
                                             & J.C.R.H. & F.L.      & S.W.J.   & D.B.H. \\ \hline
\multicolumn{1}{|l|}{conceptualization}      &equal     &supporting &supporting&equal     \\ \hline
\multicolumn{1}{|l|}{data-curation}          &equal     &equal      &          &        \\ \hline
\multicolumn{1}{|l|}{formal-analysis}        &supporting&lead      &supporting&        \\ \hline
\multicolumn{1}{|l|}{funding-acquisition}    &          &           &          &lead     \\ \hline
\multicolumn{1}{|l|}{investigation}          &lead     &supporting  &supporting&supporting\\ \hline
\multicolumn{1}{|l|}{methodology}            &equal     &equal      &equal     &equal   \\ \hline
\multicolumn{1}{|l|}{project-administration} &          &           &          &lead     \\ \hline
\multicolumn{1}{|l|}{resources}              &          &           &          &lead     \\ \hline
\multicolumn{1}{|l|}{software}               &equal     &equal      &supporting&         \\ \hline
\multicolumn{1}{|l|}{supervision}            &          &           &supporting&lead     \\ \hline
\multicolumn{1}{|l|}{validation}             &equal     &equal      &supporting&equal     \\ \hline
\multicolumn{1}{|l|}{visualization}          &equal     &equal      &supporting&supporting\\ \hline
\multicolumn{1}{|l|}{writing-original-draft} &equal     &equal      &supporting&supporting\\ \hline
\multicolumn{1}{|l|}{writing-review-editing} &equal     &equal      &supporting&equal     \\ \hline
\end{tabular}}
\end{table}

\section*{Data Availability}

The data that support the findings of this study are openly available in Zenodo at 
\href{https://doi.org/10.5281/zenodo.10657940}{https://doi.org/10.5281/zenodo.10657940}, 
reference number~\cite{rivera_hernandez_2024_10657940}.

\appendix

\section{Derivation of the theoretical model}\label{appA}

The circuit of the JPA shown in Fig.~\ref{fig:schematics}a corresponds to a lumped-element LC circuit with a non-linear Josephson inductance $L(\phi)=\frac{\phi_0}{4\pi I_c|\cos(\pi\phi/\phi_0)|}$. 
By Taylor-expanding $\frac{1}{L(\phi)}\simeq \frac{1}{L_J}[1 + \kappa_1\phi^2 + \kappa_2\phi^4]$, and considering a small time perturbation of $1/L_J$ by the pumping signal $g_p(t)$, one obtains the Hamiltonian~\cite{yamamoto_principles_2016}
\begin{equation}
    H=\frac{1}{2C}q^2 + \frac{1}{2L_0}\phi^2 + \frac{\omega_0}{2}g_p(t)\phi^2 + \kappa_1 \phi^4 
    \label{eq:H0a}
\end{equation}
where $\omega_0$ is the resonant frequency of the unperturbed oscillator, $q$ the charge in the capacitor and $\phi$ the magnetic flux in the dc-SQUID. 
Introducing the ladder operators, $A=\frac{\phi + iq}{\sqrt{2}}$ and $A^\dag=\frac{\phi - iq}{\sqrt{2}}$, one derives the quartic Hamiltonian
\begin{equation}
    H=\frac{\omega_0}{2}\left(A^\dag A + AA^\dag\right) + \frac{\omega_0}{2}g_p(t)\left[A + A^\dag\right]^2 + \kappa_1\left[A + A^\dag\right]^4 .
    \label{eq:H0a2}
\end{equation}
The quartic term in \eqref{eq:H0a2} is responsible for the Kerr non-linearity in the equation of motion~\cite{yamamoto_principles_2016}. 
In the limit of weak pumping, the quartic term can be neglected~\cite{yamamoto_principles_2016, petrovnin_generation_2023} leading to simpler linear dynamics.
In this limit, the Hamiltonian of the JPA reads
\begin{equation}
    H=\frac{\omega_0}{2}\left(A^\dag A + AA^\dag\right) + \frac{\omega_0}{2}g_p(t)\left[A + A^\dag\right]^2
    \label{eq:H0a3}
\end{equation}
Eq.~(\ref{eq:H0a3}) shows a quadratic, time-dependent Hamiltonian where the second term describes the dependence of the system's dynamics on the pumping signal $g_p(t)$.
From (\ref{eq:H0a3}) the equations of motion can be readily derived using the standard canonical approach. 
By adding the input driving contribution coupled to the oscillator by a factor $\sqrt{\gamma}$~\cite{yamamoto_principles_2016, naaman_synthesis_2022} one recovers the usual Quantum Langevin equations of motion
\begin{equation}
    \left\{\begin{array}{lcr}
        \frac{dA}{dt}&=&-i\tilde\omega_0A-i\omega_0g_p(t)(A + A^\dag) +\sqrt{\gamma}\;a_{\text{in}}\\
        \frac{dA^\dag}{dt}&=&i\tilde\omega_0^*A^\dag+i\omega_0g_p(t)(A + A^\dag) +\sqrt{\gamma}\;a^\dag_{in}
    \end{array}\right.
\label{eq:EOM0a}
\end{equation}
where $\tilde\omega_0= \omega_0 - i\frac{\gamma}{2}$, and $\gamma$ describes the losses due to the coupling to the transmission line.
In the classical limit, $A^\dag\equiv A^*$, the equations of motion become a system of two, coupled, linear equations in the complex-valued functions of time $A\equiv A(t)$ and $A^* \equiv A^*(t)$.

\begin{figure}[t]
\centering
    \includegraphics[width=\columnwidth]{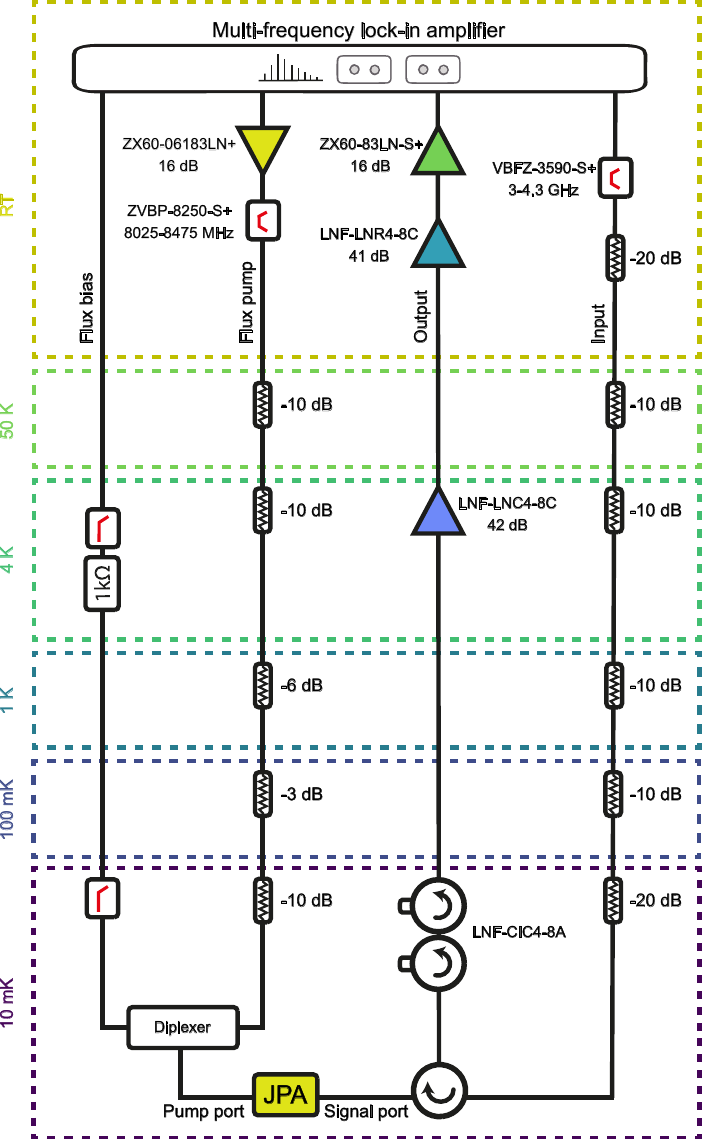}
    \caption{Schematic of the measurement setup including room temperature and cryogenic electronics.}
\label{fig:Setup} 
\end{figure}

Let us now define an orthonormal set of frequency modes $c_i(t)=e^{-i\omega_it}/\sqrt{T}$, properly defined from the time interval of the measurement $t\in[0, T]$, together with the scalar product that ensures the orthogonality condition
\begin{equation}
    \int_T dt \,c_i^*(t)c_j(t)=\delta_{ij} .
\end{equation} 
For the orthogonality condition to be satisfied, modes frequencies $\omega_i$ would need to feature a period of oscillation $T_i=2\pi/\omega_i$ commensurate to $T$ (i.e. $T=nT_i$ for $n\in \mathbb{N}$).
By expanding $A$ and $A^*$ in the frequency-mode basis
\begin{equation}
    A(t) = \sum_i a_i(t) = \sum_i a_i c_i(t), \;\;\;\; A^*(t) = \sum_i a_i^* c_i^*(t)
\end{equation}
where $a_i=\int_T dt \, c_i^*(t) A(t) \in\mathbb{C}$ are the complex amplitudes of mode $i$; the Langevin equations become
\begin{multline}
    \sum_i \Big[ \frac{da_{i}(t)}{dt}+i\tilde\omega_0 a_{i}(t) + i\omega_0 g_p(t) \left(a_{i}(t) + a^*_{i}(t)\right) \\ 
    -\sqrt{\gamma} \; a_{\text{in},\,i}(t) \Big] = 0. 
\label{eq:EOM0a-2}
\end{multline}
For the sake of brevity, we omitted the complex conjugate equation in (\ref{eq:EOM0a-2}), but it is intended to be part of the system of equations.
By Fourier transforming Eq.~(\ref{eq:EOM0a-2}), the time derivatives $\frac{d}{dt}\rightarrow i\omega$, and the system of equations of motion becomes
\begin{multline}
    \sum_i \left[i\omega a_{i}(\omega) + i\tilde\omega_0 a_{i}(\omega) + i\omega_0 \, g_p(\omega)*\left(a_{i}(\omega)+ a^*_{i}(\omega)\right) \right. \\
    \left. -\sqrt{\gamma} \; a_{\text{in},\,i}(\omega) \right]=0.
\label{EOM0a1}
\end{multline}
where $a_i(\omega)=a_i\varphi_i(\omega)$.
$\varphi_i(\omega)$ is the frequency domain representation of mode $i$, which is the Fourier transform of a cosine function of frequency $\omega_i$ multiplied by a square window of length $T$, namely a \emph{sinc} function centered around $\omega_i$. 
As shown in Fig.~\ref{fig:schematics}b, if the integer proportionality between the mode's periodicity $T_i$ and window's length $T$ is chosen, the maximum of each mode $j$ sits on the \emph{sinc} nodes of the other modes. 
This choice eliminates spectral leakage between modes and sets the mode spacing and frequency resolution of the measurement basis to $\Delta=1/T$. 
In practice we achieve these conditions by tuning the mode frequencies in relation to the sampling clock of our digital microwave platform.

The product in the time domain \eqref{eq:EOM0a-2} becomes a convolution in the frequency domain \eqref{EOM0a1}.
Let us consider the pumping signal as composed of a sum of pure frequency tones 
\begin{equation}
    g_p(t) = \sum_k A_k cos\left(\Omega_k t + \phi_k\right) = \sum_k g_k e^{i\Omega_kt} + g_k^* e^{-i\Omega_kt}
\label{gpt}
\end{equation}
where $g_k=\frac{1}{2}A_k e^{i\phi_k} \in \mathbb{C}$ represents the complex pump strength. 
The Fourier transform of Eq.~(\ref{gpt}) is
\begin{equation}
    g_p(\omega) = \sum_k g_k \delta(\omega-\Omega_k) + g_k^*\delta(\omega+\Omega_k).
\end{equation}
Exploiting the convolution properties of the Dirac's deltas $a_i(\omega)*\delta(\omega \pm \Omega_k) = a_i \varphi_i (\omega \pm \Omega_k)$ we eliminate the convolution 
in Eq.~(\ref{EOM0a1}). 
If $\Omega_k$ and $\omega_i$ are commensurate such that $\Omega_k+\omega_i=\omega_j$, that is, $a_i(\omega)*\delta(\omega\pm\Omega_k)=a_i\varphi_j(\omega)$, one obtains mode index exchange~\cite{petrovnin_generation_2023}. 
This implies the appearance of mixed terms $a_i\varphi_j(\omega)$ with $i\neq j$ in the equations of motion. 
However, by factoring out of the sum on $p$ the terms with the same index $j$, one can rewrite the equations of motion as
\begin{multline}
    \sum_i \left[(i\omega + i\tilde\omega_0) a_{i} \varphi_{i}(\omega) + i\sum_j \left(g_{ij} a_{i} + g_{ij}^\prime a_{i}^* \right) \varphi_{j}(\omega) \right.\\ 
    \Bigg. -\sqrt{\gamma} \; a_{\text{in},\,i} \varphi_{i} (\omega)\Bigg] = 0.
\label{EOM0a2}
\end{multline}
where $g_{ij}$ ($g_{ij}^\prime$) are the couplings between $a_i$ ($a_i^*$) and $\varphi_j$.
Depending on the pump frequencies $\Omega_k$ connecting mode $i$ to mode $j$, the coupling takes the form $g_{ij}=\omega_0 g_k$ ($g_{ij}=\omega_0 g_k^*$) if connected by $+\Omega_k$ ($-\Omega_k$). 
Notice that couplings $g_{ij}$ and $g_{ij}^\prime$ may differ, depending on the particular pumping scheme adopted.
Writing explicitly the two sums in Eq.~(\ref{EOM0a2}) and grouping together terms sharing the same $\varphi_{j}(\omega)$, one can rewrite Eq.~(\ref{EOM0a2}) as
\begin{multline}
    \sum_m \left[ (i\omega +i\tilde\omega_0) a_{m} + i\sum_\ell \left(g_{\ell m} a_{\ell}+ g_{\ell m}^\prime a_{\ell}^* \right) \right.\\
    \Bigg. -\sqrt{\gamma} \; a_{\text{in},\,m}\Bigg] \varphi_{m}(\omega) = 0. 
\label{EOM0a3}
\end{multline}
Equation~\eqref{EOM0a3} is solved if each term in square brackets is zero.
Integrating over $\omega$ we recover the system of $2N$-coupled linear equations Eq.~\eqref{EOM0}, where $N$ is the number of measured modes.
\begin{equation}
    \left\{\begin{array}{rcl}
        i\omega_1 a_1 + i\tilde\omega_0 a_1 +i\sum_j g_{1j} (a_j + a_j^*) &=& \sqrt{\gamma} \; a_{\text{in},\,1} \\
        -i\omega_1 a_1^* - i\tilde\omega_0^* a_1^* -i\sum_j g_{1j} (a_j + a_j^*) &=&\sqrt{\gamma}\;a_{\text{in},\,1}^*\\
        i\omega_2 a_2 + i\tilde\omega_0 a_2 +i\sum_j g_{2j} (a_j + a_j^*) &=& \sqrt{\gamma}\;a_{\text{in},\,2} \\
        -i\omega_2 a_2^* - i\tilde\omega_0^* a_2^* -i\sum_j g_{2j} (a_j + a_j^*) &=& \sqrt{\gamma} \; a_{\text{in},\,2}^* \\
        &\vdots& \\
        i\omega_i a_i + i\tilde\omega_0 a_i +i\sum_j g_{ij} (a_j + a_j^*) &=& \sqrt{\gamma}\;a_{\text{in},\,i} \\
        -i\omega_i a_i^* - i\tilde\omega_0^* a_i^* -i\sum_j g_{ij} (a_j + a_j^*) &=& \sqrt{\gamma} \; a_{\text{in},\,i}^*
    \end{array} \right.
\label{nEOMa}
\end{equation}
In Eq.~(\ref{nEOMa}) we take $g_{ij}=g_{ij}^\prime$ for simplicity, but this is not the general case.
When integrating over $\omega$ we use the fact that in the limit $\Delta \rightarrow 0$ integrals $\int \omega\varphi_j(\omega)\,d\omega\approx\int\omega\delta(\omega-\omega_j)\,d\omega=\omega_j$ and $\int \varphi_j(\omega)\,d\omega=1$.

\begin{figure}[t]
\centering
\includegraphics[width=\columnwidth]{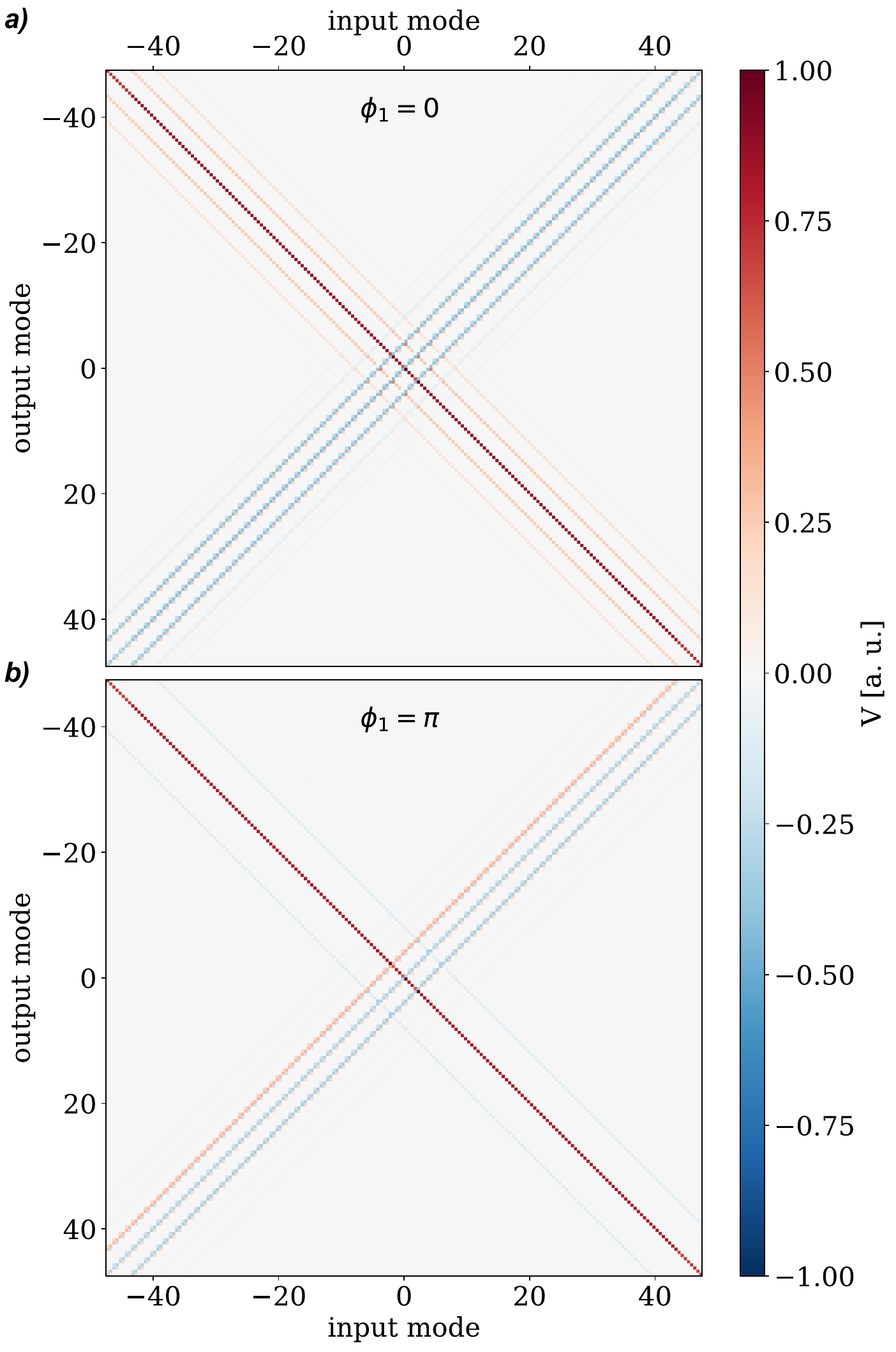}
\caption{Simulated covariance matrix of the triply pumped JPA for phase a) $\phi_1=0$ and b) $\phi_1=\pi$.
The simulation transforms an input covariance matrix through equation~\eqref{S_transform} based on the scattering matrices of Fig.~\ref{fig:S_3pmp}.
The input covariance matrix is obtained by simulating $2n$ Gaussian uncorrelated random processes with zero-mean and standard deviation $1/\sqrt{2}$ (arbitrary units).}
\label{fig:V_sim}
\end{figure}

A standard convention in literature is to derive the above set of equations within the Rotating Wave Approximation (RWA)~\cite{peterson_parametric_2020, naaman_synthesis_2022, petrovnin_generation_2023}. 
The time dependence of the pump is eliminated by moving to a reference frame co-rotating with the pump (redefining the ladder operators $A \rightarrow A\, e^{\,i\frac{\Omega_k}{2}t}$)~\cite{yamamoto_principles_2016}. 
While this is perfectly valid in the case of a single pump, it becomes more difficult to interpret when multiple pump frequencies are introduced. 
We prefer the approach adopted above relying on the convolution properties of delta functions, without the need to introduce a change of reference frame.

\begin{figure}[t]
\centering
\includegraphics[width=\columnwidth]{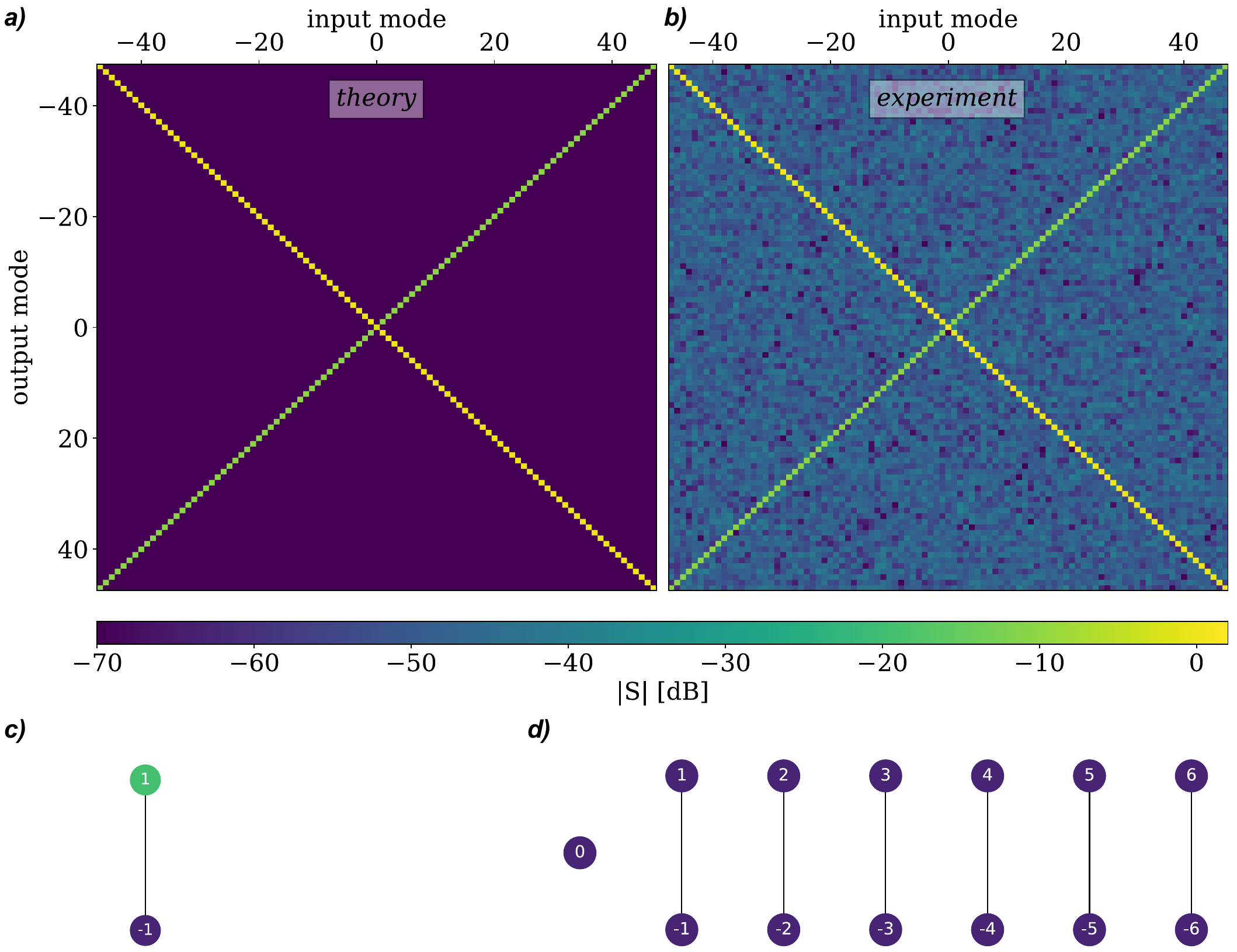}
\caption{Theoretical a) and experimental b) scattering matrices for the case of one parametric pump at $\Omega_0\approx2\omega_0$. Panels c) and d) show the single-mode correlation and total correlation graphs respectively.}
\label{fig:S_1pmp} 
\end{figure}

\subsection{Definition of the Scattering Matrix}

We write the linear system of equations (\ref{nEOMa}) in the matrix form following the standard formalism of mode scattering theory \cite{peterson_parametric_2020, naaman_synthesis_2022, jolin_multipartite_2023, petrovnin_generation_2023}
\begin{equation}
    -iM\vec{a}=K\vec{a}_{in}.
    \label{m_form}
\end{equation}
The solutions are readily available upon inversion of matrix $M$. 
The mode-coupling matrix $K$ is a diagonal matrix containing the wave-guide couplings for each mode (assumed to be constant $K_{ii}=\sqrt{\gamma}$ $\forall i$), while $\vec{a}$ and $\vec{a}_{in}$ are the vectors containing complex amplitudes, e.g. $\vec{a}=(a_1, a_1^*, a_2, a_2^*, \dots, a_i, a_i^*, \dots)^T$.

The scattering matrix $S$ is defined as the input-output relationship
\begin{equation}
    \vec{a}_{out}=S\cdot\vec{a}_{in}
\label{Sdef}
\end{equation}
By considering the mode matching condition $\vec{a}_{in} + \vec{a}_{out}=K\vec{a}$ between the mode basis internal to the JPA and the external input/output modes of the transmission line~\cite{peterson_parametric_2020, naaman_synthesis_2022}, one retrieves the usual definition of the scattering matrix:
\begin{equation}
S=iKM^{-1}K-I.
\label{S_app}
\end{equation}

\subsection{Covariance matrix and correlations of Gaussian states} \label{app-covmat}

Correlations in a Gaussian state are determined by the covariance matrix $V$. 
Subject to a Gaussian Unitary transformation described by the real symplectic matrix $S_x$,  $V$ transforms as
\begin{equation}
    V_{out}= S_{x}V_{in}S_{x}^T.
    \label{S_transform}
\end{equation}
The quadratic form of the Hamiltonian describing the weakly-pumped JPA defines the matrix $S_x = U S U^\dag$ which is simply the scattering matrix expressed in the multimodal quadratures $x_i,\,p_i$ basis.  
The transformation $U$ corresponds to the usual canonical transformation between modes quadratures and creation and annihilation operators $x_i=\frac{a_i + a^*_i}{\sqrt{2}}$ and $p_i=\frac{a_i - a^*_i}{\sqrt{2}i}$.
For the experiments considered in this manuscript $V_{in}$ takes the form of a $2n \times 2n$ diagonal matrix corresponding to $n$ uncorrelated Gaussian modes input into the JPA.
For a complete review of Gaussian quantum resources, we refer to \cite{weedbrook_gaussian_2012}.
It follows from (\ref{S_transform}) that if $V_{in} \propto \mathbb{I}$, $V_{out}$ is solely determined by $S_x$.
When the scattering matrices $S$ and covariance matrix $V$ have the same mode connectivity, $S_{ij}\neq 0$ implies $V_{ij}\neq 0$.  
Here $V_{ij}$ is a $2\times2$ matrix containing the correlations $\langle \hat{q}_i\hat{q}_j\rangle$ with $\hat{q}\in\{\hat{x},\hat{p}\}$ between the quadratures of any pair of modes $i$~and~$j$.

As an example, we use~\eqref{S_transform} to simulate what would happen when vacuum fluctuations are input into the weakly-pumped JPA. 
The vacuum state $\ket{0}$ is a Gaussian state with covariance matrix $V_{in}=\frac{\hbar}{2}\mathbb{I}$. 
We expect to obtain an output covariance matrix $V_{out}$ having the same topology of correlations described by the scattering matrix $S$.
In Fig.~\ref{fig:V_sim} we simulated the transformation of the vacuum through equation \eqref{S_transform}, where the input vacuum was obtained by simulating $2n$ Gaussian independent random processes.
The simulation is carried out for both values of the phase $\phi_1$ of the $xp$ representation of the scattering matrix of Fig.~\ref{fig:S_3pmp}.
We can see that for both phases the output covariance preserves the same mode connectivity described by $S$.

\subsection{Fitting experimental and theoretical $S$ matrices}\label{app-fit}

\begin{figure}[t]
\centering
\includegraphics[width=\columnwidth]{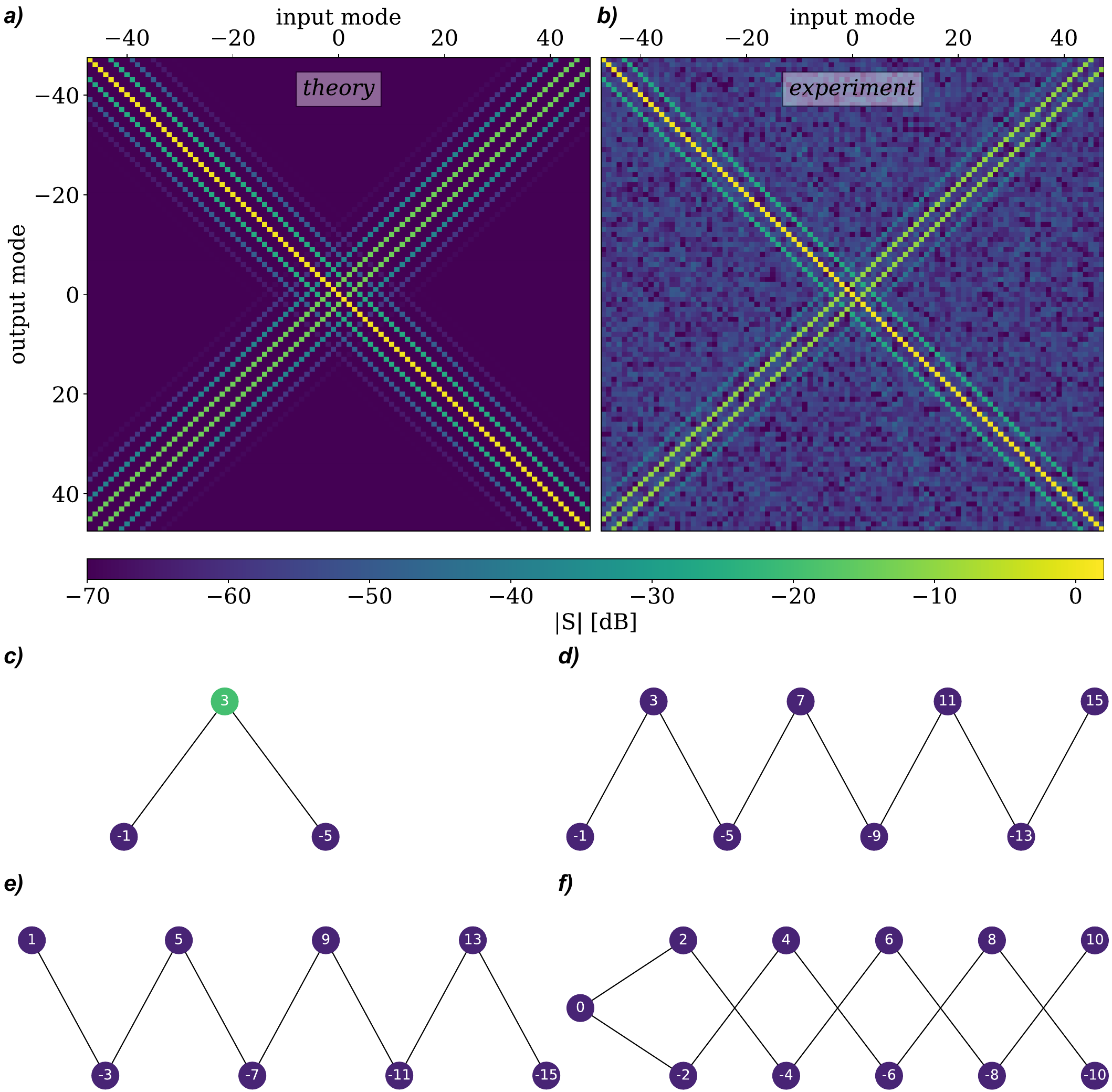}
\caption{Theoretical a) and experimental b) scattering matrices for the case of two parametric pumps at $\Omega_{k} \approx 2\omega_0 + 2k\Delta$, with $k=\{-1, 1\}$. Panel c) shows the single-mode correlation graph obtained by setting a threshold at $-20~\text{dB}$. Panels d)-f) show the three connected subgraphs that arise in the global correlation scheme.}
\label{fig:S_2pmp}
\end{figure}

To compare the numerical solution of the theoretical model with the experimental results, we normalize both theory and experimental data to the respective zero-pump case. 
Throughout this work we considered the balanced case where all the pump signals are tuned to the same strength $|g_{-1}|=|g_{0}|=|g_{1}|=g$.
To compare the theoretical model \eqref{nEOMa} with our experiment, it is sufficient to treat $g$ and $\gamma$ as free parameters and fit their values. 
The fit is performed by solving Eq.~(\ref{nEOMa}) for different values of $g$ and $\gamma$ to determine a theoretical $S_{T}(g,\gamma)$, which is compared to the experimental scattering matrix $S_{E}$.  
By minimizing the distance function
\begin{equation}
    d(g,\gamma)=\sqrt{\sum_{ij}| S_{E,\;ij} - S_{T}(g,\gamma)_{ij} |^2},
\label{eq:dist}
\end{equation}
we find the best-fit $g$ and $\gamma$.
For three parametric pumps, we found that the minima of \eqref{eq:dist} lie along a line $\omega_0 g/\gamma\approx 0.14$.
Our assumption of frequency-independent coupling, a valid approximation when all modes are well within the bandwidth of the JPA, makes the fitting procedure sensitive to only the ratio $g/\gamma$.

\section{Measurement setup}\label{app-setup}

A schematic of the measurement setup is shown in Fig.~\ref{fig:Setup}.
The JPA is cooled at $10~\text{mk}$ in a Bluefors LD250 dry dilution refrigerator with appropriate attenuation of the pump and signal at in multiple stages.
We operate the JPA in reflection with a circulator to separate the input and output fields.
The measurement signal-to-noise ratio is set by the first High-Electron-Mobility Transistor (HEMT) amplifier at $4~\text{K}$.
A double isolator reduces the amplifier noise backacting on the sample.

A multi-frequency lock-in amplifier \cite{tholen_measurement_2022} provides the probe signal and the multi-frequency pump through separate output ports which are synchronous with the input measurement port.
It also provides the DC bias for the JPA which is combined with the pump in a diplexer at base temperature.
The lockin uses direct digital synthesis to create the signal in the band $3-6~\text{GHz}$ and the multi-frequency pump in the band $5-10~\text{GHz}$ with high-speed digital-to-analog converters working in the 2nd Nyquist zone.  
Similarly, input signals appearing in a higher Nyquist zone of the analog-to-digital converter are under-sampled, appearing at aliased frequencies in the baseband.
Analog filters at room temperature select the appropriate Nyquist zone.
In this way we perform digital signal processing at microwave frequencies without analog IQ mixers.
This direct digital approach is essential for maintaining the orthogonality of the mode basis and eliminating Fourier leakage between modes.

\section{One and two parametric pumps results}\label{App12pmp}

As a prequel to the more complex three-frequency pump, it is important to understand and verify the more simple single and double pump case.
We use the same methodology explained in section~\ref{exp} and the same experimental setup.
Results shown here for one and two pumps are consistent with the covariance matrix analysis published in previous work~\cite{jolin_multipartite_2023}.

A single monochromatic pump applied at roughly twice the resonant frequency of the JPA $\Omega_0 \approx2 \omega_0$ enables 2nd order intermodulation between the pump and signal, generating intermodulation products (idlers)  $\omega_j = \Omega_0/2 + j\Delta$, with $j\in\mathbb{Z}$ (three-wave mixing).
The scattering matrix for the single pumped JPA is shown in Fig.~\ref{fig:S_1pmp} where we see the anti-diagonal mixing products.
From the scattering matrix we also generate the single-mode correlation graph shown in Fig.~\ref{fig:S_1pmp}c.
There are only pair-wise connections in the generalized graph Fig.~\ref{fig:S_1pmp}d.
We verify that the scattering matrix is independent of the phase of the pump.

The frequencies of the bichromatic pump are roughly centered at twice the JPA resonance and detuned by twice the spacing of the mode basis, $\Omega_{k} = 2\omega_0 + 2k\Delta$, with $k=\{-1, 1\}$.
Our mode basis becomes $\omega_j = \frac{1}{4} \left(\Omega_{-1} + \Omega_1 \right) + j \Delta$, with $j\in\mathbb{Z}$.
In Fig.~\ref{fig:S_2pmp} we show the scattering matrix for the double-pumped JPA.
This pumping scheme creates single-mode correlations shown in Fig.~\ref{fig:S_2pmp}c), where each mode is connected to two mixing products.
Generalizing the single-mode graph over all modes in the scattering matrix, we find three independent linear subgraphs shown in Fig.~\ref{fig:S_2pmp}d-e.
We verify that the scattering matrix is independent of the relative phase of the two pumps. 
Experimental results, in both cases, show good agreement with our theoretical predictions.

\bibliographystyle{unsrt}
\bibliography{Refs}     

\begin{thebibliography}{10}

\bibitem{grimm_stabilization_2020}
A.~Grimm, N.~E. Frattini, S.~Puri, S.~O. Mundhada, S.~Touzard, M.~Mirrahimi,
  S.~M. Girvin, S.~Shankar, and M.~H. Devoret.
\newblock Stabilization and operation of a kerr-cat qubit.
\newblock {\em Nature}, 584(7820):205--209, 2020.

\bibitem{kudra_robust_2022}
Marina Kudra, Mikael Kervinen, Ingrid Strandberg, Shahnawaz Ahmed, Marco
  Scigliuzzo, Amr Osman, Daniel~Pérez Lozano, Mats~O. Tholén, Riccardo
  Borgani, David~B. Haviland, Giulia Ferrini, Jonas Bylander, Anton~Frisk
  Kockum, Fernando Quijandría, Per Delsing, and Simone Gasparinetti.
\newblock Robust preparation of wigner-negative states with optimized
  {SNAP}-displacement sequences.
\newblock {\em {PRX} Quantum}, 3(3):030301, 2022.

\bibitem{sivak_real-time_2023}
V.~V. Sivak, A.~Eickbusch, B.~Royer, S.~Singh, I.~Tsioutsios, S.~Ganjam,
  A.~Miano, B.~L. Brock, A.~Z. Ding, L.~Frunzio, S.~M. Girvin, R.~J.
  Schoelkopf, and M.~H. Devoret.
\newblock Real-time quantum error correction beyond break-even.
\newblock {\em Nature}, 616(7955):50--55, 2023.

\bibitem{braunstein_quantum_2005}
Samuel~L. Braunstein and Peter van Loock.
\newblock Quantum information with continuous variables.
\newblock {\em Rev. Mod. Phys.}, 77(2):513--577, 2005.

\bibitem{weedbrook_gaussian_2012}
Christian Weedbrook, Stefano Pirandola, Raúl García-Patrón, Nicolas~J. Cerf,
  Timothy~C. Ralph, Jeffrey~H. Shapiro, and Seth Lloyd.
\newblock Gaussian quantum information.
\newblock {\em Rev. Mod. Phys.}, 84(2):621--669, 2012.

\bibitem{pfister_continuous-variable_2019}
Olivier Pfister.
\newblock Continuous-variable quantum computing in the quantum optical
  frequency comb.
\newblock {\em J. Phys. B: At. Mol. Opt. Phys.}, 53(1):012001, 2019.

\bibitem{menicucci_universal_2006}
Nicolas~C. Menicucci, Peter van Loock, Mile Gu, Christian Weedbrook, Timothy~C.
  Ralph, and Michael~A. Nielsen.
\newblock Universal quantum computation with continuous-variable cluster
  states.
\newblock {\em Physical Review Letters}, 97(11):110501, 2006.

\bibitem{menicucci_one-way_2008}
Nicolas~C. Menicucci, Steven~T. Flammia, and Olivier Pfister.
\newblock One-way quantum computing in the optical frequency comb.
\newblock {\em Phys. Rev. Lett.}, 101(13):130501, 2008.

\bibitem{gu_quantum_2009}
Mile Gu, Christian Weedbrook, Nicolas~C. Menicucci, Timothy~C. Ralph, and Peter
  van Loock.
\newblock Quantum computing with continuous-variable clusters.
\newblock {\em Physical Review A}, 79(6):062318, 2009.

\bibitem{menicucci_graphical_2011}
Nicolas~C. Menicucci, Steven~T. Flammia, and Peter van Loock.
\newblock Graphical calculus for gaussian pure states.
\newblock {\em Phys. Rev. A}, 83(4):042335, 2011.

\bibitem{pfister_multipartite_2004}
Olivier Pfister, Sheng Feng, Gregory Jennings, Raphael Pooser, and Daruo Xie.
\newblock Multipartite continuous-variable entanglement from concurrent
  nonlinearities.
\newblock {\em Phys. Rev. A}, 70(2):020302, 2004.

\bibitem{bensemhoun_multipartite_2023}
Adrien Bensemhoun, C.~Gonzalez-Arciniegas, Olivier Pfister, Laurent Labont,
  Jean Etesse, Anthony Martin, Sbastien Tanzilli, Giuseppe Patera, and Virginia
  d'Auria.
\newblock Multipartite entanglement in bright frequency combs from
  microresonators, 2023.

\bibitem{yang_squeezed_2021}
Zijiao Yang, Mandana Jahanbozorgi, Dongin Jeong, Shuman Sun, Olivier Pfister,
  Hansuek Lee, and Xu~Yi.
\newblock A squeezed quantum microcomb on a chip.
\newblock {\em Nature Communications}, 12(1):4781, 2021.

\bibitem{jahanbozorgi_spectroscopic_2023}
Mandana Jahanbozorgi, Zijiao Yang, Emily~A. Parnell, Dongin Jeong, Shuman Sun,
  Olivier Pfister, Hansuek Lee, and Xu~Yi.
\newblock Spectroscopic characterization of an on-chip squeezed quantum optical
  frequency comb.
\newblock In {\em {CLEO} 2023 (2023), paper {FF}1L.2}, page FF1L.2. Optica
  Publishing Group, 2023.

\bibitem{kouadou_spectrally_2023}
Tiphaine Kouadou, F.~Sansavini, M.~Ansquer, J.~Henaff, N.~Treps, and V.~Parigi.
\newblock Spectrally shaped and pulse-by-pulse multiplexed multimode squeezed
  states of light.
\newblock {\em {APL} Photonics}, 8(8):086113, 2023.

\bibitem{menicucci_arbitrarily_2010}
Nicolas~C. Menicucci, Xian Ma, and Timothy~C. Ralph.
\newblock Arbitrarily large continuous-variable cluster states from a single
  quantum nondemolition gate.
\newblock {\em Phys. Rev. Lett.}, 104(25):250503, 2010.

\bibitem{pysher_parallel_2011}
Matthew Pysher, Yoshichika Miwa, Reihaneh Shahrokhshahi, Russell Bloomer, and
  Olivier Pfister.
\newblock Parallel generation of quadripartite cluster entanglement in the
  optical frequency comb.
\newblock {\em Phys. Rev. Lett.}, 107(3):030505, 2011.

\bibitem{cai_multimode_2017}
Y.~Cai, J.~Roslund, G.~Ferrini, F.~Arzani, X.~Xu, C.~Fabre, and N.~Treps.
\newblock Multimode entanglement in reconfigurable graph states using optical
  frequency combs.
\newblock {\em Nat Commun}, 8(1):15645, 2017.

\bibitem{walschaers_tailoring_2018}
Mattia Walschaers, Supratik Sarkar, Valentina Parigi, and Nicolas Treps.
\newblock Tailoring non-gaussian continuous-variable graph states.
\newblock {\em Phys. Rev. Lett.}, 121(22):220501, 2018.

\bibitem{menicucci_temporal-mode_2011}
Nicolas~C. Menicucci.
\newblock Temporal-mode continuous-variable cluster states using linear optics.
\newblock {\em Phys. Rev. A}, 83(6):062314, 2011.

\bibitem{chen_experimental_2014}
Moran Chen, Nicolas~C. Menicucci, and Olivier Pfister.
\newblock Experimental realization of multipartite entanglement of 60 modes of
  a quantum optical frequency comb.
\newblock {\em Physical Review Letters}, 112(12):120505, 2014.

\bibitem{asavanant_time-domain-multiplexed_2021}
Warit Asavanant, Baramee Charoensombutamon, Shota Yokoyama, Takeru Ebihara,
  Tomohiro Nakamura, Rafael~N. Alexander, Mamoru Endo, Jun-ichi Yoshikawa,
  Nicolas~C. Menicucci, Hidehiro Yonezawa, and Akira Furusawa.
\newblock Time-domain-multiplexed measurement-based quantum operations with
  25-{MHz} clock frequency.
\newblock {\em Phys. Rev. Appl.}, 16(3):034005, 2021.

\bibitem{du_generation_2023}
Peilin Du, Yu~Wang, Kui Liu, Rongguo Yang, and Jing Zhang.
\newblock Generation of large-scale continuous-variable cluster states
  multiplexed both in time and frequency domains.
\newblock {\em Opt. Express, {OE}}, 31(5):7535--7544, 2023.

\bibitem{larsen_deterministic_2019}
Mikkel~V. Larsen, Xueshi Guo, Casper~R. Breum, Jonas~S. Neergaard-Nielsen, and
  Ulrik~L. Andersen.
\newblock Deterministic generation of a two-dimensional cluster state.
\newblock {\em Science}, 366(6463):369--372, 2019.

\bibitem{jing_experimental_2006}
Jietai Jing, Sheng Feng, Russell Bloomer, and Olivier Pfister.
\newblock Experimental continuous-variable entanglement from a
  phase-difference-locked optical parametric oscillator.
\newblock {\em Phys. Rev. A}, 74(4):041804, 2006.

\bibitem{hung_quantum_2021}
Jimmy~S.C. Hung, J.H. Busnaina, C.W.~Sandbo Chang, A.M. Vadiraj, I.~Nsanzineza,
  E.~Solano, H.~Alaeian, E.~Rico, and C.M. Wilson.
\newblock Quantum simulation of the bosonic creutz ladder with a parametric
  cavity.
\newblock {\em Phys. Rev. Lett.}, 127(10):100503, 2021.

\bibitem{petrovnin_generation_2023}
Kirill~Viktorovich Petrovnin, Michael~Romanovich Perelshtein, Tero Korkalainen,
  Visa Vesterinen, Ilari Lilja, Gheorghe~Sorin Paraoanu, and Pertti~Juhani
  Hakonen.
\newblock Generation and structuring of multipartite entanglement in a
  josephson parametric system.
\newblock {\em Advanced Quantum Technologies}, 6(1):2200031, 2023.

\bibitem{pfister_cluster_2023}
Olivier Pfister, Carlos Gonzalez-Arciniegas, Xuan Zhu, Chun-Hung Chang, and Avi
  Pe'er.
\newblock Cluster quantum state generation based on phase modulated optical
  parametric oscillator, 2023.

\bibitem{andersson_squeezing_2022}
Gustav Andersson, Shan~W. Jolin, Marco Scigliuzzo, Riccardo Borgani, Mats~O.
  Tholén, J.C. Rivera~Hernández, Vitaly Shumeiko, David~B. Haviland, and Per
  Delsing.
\newblock Squeezing and multimode entanglement of surface acoustic wave
  phonons.
\newblock {\em {PRX} Quantum}, 3(1):010312, 2022.

\bibitem{naaman_synthesis_2022}
Ofer Naaman and José Aumentado.
\newblock Synthesis of parametrically coupled networks.
\newblock {\em {PRX} Quantum}, 3(2):020201, 2022.

\bibitem{petrovnin_microwave_2023}
Kirill Petrovnin, Jiaming Wang, Michael Perelshtein, Pertti Hakonen, and
  Gheorghe~Sorin Paraoanu.
\newblock Microwave photon detection at parametric criticality, 2023.

\bibitem{esposito_observation_2022}
Martina Esposito, Arpit Ranadive, Luca Planat, Sébastien Leger, Dorian
  Fraudet, Vincent Jouanny, Olivier Buisson, Wiebke Guichard, Cécile Naud,
  José Aumentado, Florent Lecocq, and Nicolas Roch.
\newblock Observation of two-mode squeezing in a traveling wave parametric
  amplifier.
\newblock {\em Phys. Rev. Lett.}, 128(15):153603, 2022.

\bibitem{jolin_multipartite_2023}
Shan~W. Jolin, Gustav Andersson, J. C.~Rivera Hernández, Ingrid Strandberg,
  Fernando Quijandría, José Aumentado, Riccardo Borgani, Mats~O. Tholén, and
  David~B. Haviland.
\newblock Multipartite entanglement in a microwave frequency comb.
\newblock {\em Phys. Rev. Lett.}, 130(12):120601, 2023.

\bibitem{ranzani_graph-based_2015}
Leonardo Ranzani and José Aumentado.
\newblock Graph-based analysis of nonreciprocity in coupled-mode systems.
\newblock {\em New J. Phys.}, 17(2):023024, 2015.

\bibitem{serafini_quantum_2017}
Alessio Serafini.
\newblock {\em Quantum Continuous Variables: A Primer of Theoretical Methods}.
\newblock {CRC} Press, 2017.

\bibitem{peterson_parametric_2020}
Gabriel~Aaron Peterson.
\newblock {\em Parametric Coupling between Microwaves and Motion in Quantum
  Circuits: Fundamental Limits and Applications}.
\newblock PhD thesis, University of Colorado at Boulder, 2020.

\bibitem{yamamoto_principles_2016}
Yoshihisa Yamamoto and Kouichi Semba.
\newblock {\em Principles and Methods of Quantum Information Technologies},
  volume 911 of {\em Lecture Notes in Physics}.
\newblock Springer Japan, 2016.

\bibitem{tholen_measurement_2022}
Mats~O. Tholén, Riccardo Borgani, Giuseppe~Ruggero Di~Carlo, Andreas
  Bengtsson, Christian Križan, Marina Kudra, Giovanna Tancredi, Jonas
  Bylander, Per Delsing, Simone Gasparinetti, and David~B. Haviland.
\newblock Measurement and control of a superconducting quantum processor with a
  fully integrated radio-frequency system on a chip.
\newblock {\em Review of Scientific Instruments}, 93(10):104711, 2022.

\bibitem{rivera_hernandez_2024_10657940}
J.~C. Rivera~Hernández, Fabio Lingua, Shan~W. Jolin, and David~B. Haviland.
\newblock {Data Repository for the article: "Control of multi-modal scattering
  in a microwave frequency comb"}, 2024.
\newblock Available at
  \href{https://doi.org/10.5281/zenodo.10657940}{https://doi.org/10.5281/zenodo.10657940}.

\end{thebibliography}

\end{document}